\def\rn{}
\def\nn#1 #2{#2. #1}                            
\def\nnn#1 #2 #3{#2. #3. #1}                    
\def\nnnn#1 #2 #3 #4{#2. #3. #4 #1}             
\def\nnnnn#1 #2 #3 #4 #5{#2. #3. #4 #5. #1}     

\def\rf#1;#2;#3;#4;#5 {{\frenchspacing\par\rn#1, #3 {\bf #4}, #5 (#2). \par}}
\def\rrf#1;#2;#3;#4;#5 {{\frenchspacing\rn#1, #3 {\bf #4}, #5 (#2);~}}
\def\rrrf#1;#2;#3;#4;#5 {{\frenchspacing\rn#1, #3 {\bf #4}, #5 (#2).}}
\def\rg#1;#2;#3;#4;#5;#6 {{\frenchspacing\par\rn#1, #3 {\bf #4}, #5 (#2). \par}}
\def\rfbook#1;#2;#3;#4;#5 {{\frenchspacing\par\rn#1, {\it #3} (#5, #4, #2).\par}}
\def\rfprep#1;#2;#3 {{\par\frenchspacing\rn#1, #3 (#2).\par}}
\def\rrfprep#1;#2;#3 {{\frenchspacing\rn#1, #3 (#2);~}}
\def\rrrfprep#1;#2;#3 {{\frenchspacing\rn#1, #3 (#2).}}
\def\rfproc#1;#2;#3;#4;#5;#6 {{\frenchspacing\par\rn#1 #2, in {\it #3}, ed. #4 (#5: #6)\par}}
\def\rfprocp#1;#2;#3;#4;#5;#6;#7 {{\frenchspacing\par\rn#1 #2, in {\it #3}, ed. #4 (#5: #6), p#7\par}}

\def\rg#1;#2;#3;#4;#5;#6 {\par\rn#1 #2, {\it #3}, {\bf #4}, #5 (``#6'') \par}
\def\rf#1;#2;#3;#4;#5 {\par\rn#1, {\it #3}, {\bf #4}, #5 (#2)\par}
\def\rfbook#1;#2;#3;#4;#5 {{\frenchspacing\par\rn#1, {\it #3} (#4: #5, #2)\par}}
\def\rfproc#1;#2;#3;#4;#5;#6 {{\frenchspacing\par\rn#1 #2, in {\it #3}, ed. #4 (#5: #6)\par}}
\def\rfprocp#1;#2;#3;#4;#5;#6;#7 {{\frenchspacing\par\rn#1 #2, in {\it #3}, ed. #4 (#5: #6), p#7\par}}
\def\rfprep#1;#2;#3  {{\par\rn#1, #3, #2\par}}
\def\rfprepp#1;#2;#3 {{\par\rn#1 #2, #3\par}}

\def\ie{{\frenchspacing\it i.e. }}
\def\eg{{\frenchspacing\it e.g. }}

\def\beq#1{\begin{equation}\label{#1}}
\def\eeq{\end{equation}}
\def\beqa#1{\begin{eqnarray}\label{#1}}
\def\eeqa{\end{eqnarray}}

\def\spose#1{\hbox to 0pt{#1\hss}}
\def\simlt{\mathrel{\spose{\lower 3pt\hbox{$\mathchar"218$}}
     \raise 2.0pt\hbox{$\mathchar"13C$}}}
\def\simgt{\mathrel{\spose{\lower 3pt\hbox{$\mathchar"218$}}
     \raise 2.0pt\hbox{$\mathchar"13E$}}}
\def\simpropto{\mathrel{\spose{\lower 3pt\hbox{$\mathchar"218$}}
     \raise 2.0pt\hbox{$\propto$}}}

\def\ed{\end{document}}

\def\beq#1{\begin{equation}\label{#1}}
\def\eeq{\end{equation}}
\def\beqa#1{\begin{eqnarray}\label{#1}}
\def\eeqa{\end{eqnarray}}

\documentclass[twocolumn,amsmath,nofootinbib]{revtex4} 
\usepackage{graphicx}
\usepackage{dcolumn}
\usepackage{bm}
\usepackage{url}
\usepackage{endnotes}

\newcommand{\no}{\noindent}
\newcommand{\EQ}{\begin{equation}}
\newcommand{\EQA}{\begin{eqnarray}}
\newcommand{\eqa}{\end{eqnarray}}

\newcommand{\AR}{\renewcommand {\arraystretch}{1.5}
\begin{array}{l}}
\newcommand{\bAR}{\renewcommand {\arraystretch}{2}
\begin{array}{l}}
\newcommand{\ARc}{\renewcommand {\arraystretch}{1.5}
\begin{array}{c}}
\newcommand{\bARc}{\renewcommand {\arraystretch}{2}
\begin{array}{c}}
\newcommand{\ar}{\end{array} \renewcommand {\arraystretch}{1}}

\begin{document}
\input{epsf.sty}
\def\affilmrk#1{$^{#1}$}
\def\affilmk#1#2{$^{#1}$#2}

\title{A test of the CPL parameterization for rapid dark energy equation of state transitions}
\renewcommand{\theendnote}{\fnsymbol{footnote}} 
\author{Sebastian Linden and Jean-Marc Virey}
\affiliation{
\parshape 1 -3cm 24cm
\affilmk{}{Centre de Physique Th\'eorique\footnote{`Centre de Physique Th\'eorique'' is UMR 6207 - ``Unit\'e Mixte
de Recherche'' of CNRS and of the Universities ``de Provence'',
``de la M\'editerran\'ee'' and ``du Sud Toulon-Var''- Laboratory
affiliated to FRUMAM (FR 2291). Preprint CPT\_P009\_2008.}, CNRS-Luminy
Case 907, F-13288 Marseille Cedex 9, France and Universit\'e de Provence \\ }
}
\renewcommand{\thefootnote}{\arabic{footnote}} 

\begin{abstract}
We test the robustness and flexibility of the Chevallier-Polarski-Linder (CPL) parameterization of the Dark Energy equation of state
$w(z)=w_0+w_a \frac{z}{1+z}$ in recovering a four-parameter step-like fiducial model. We constrain the parameter space region of the underlying fiducial model where the CPL parameterization offers a reliable reconstruction. It turns out that non negligible biases leak into the results for recent ($z<2.5$) rapid transitions, but that CPL yields a good reconstruction in all other cases. The presented analysis is performed with supernova Ia data as forecasted for a space mission like SNAP/JDEM, combined with future expectations for the CMB shift parameter $R$ and the BAO parameter $A$.
\end{abstract}

\maketitle

\section{Introduction}

Current studies to extract the properties of a dark energy component of the universe from observational 
data focus on the determination of its equation of state $w(z)$ 
(see \eg \cite{DEconstraints}), which is the ratio of the dark energy's pressure 
to its energy density $w(z):=\frac{p_{DE}(z)}{\rho_{DE}(z)}$. Chevallier 
and Polarski \cite{Polarskiparam} and Linder
\cite{Linderparam} 
proposed the following parameterization of the equation of state:
\begin{equation}
\label{CPL}
w^{CPL}(z)=w_0+w_a \frac{z}{1+z},
\end{equation}
hereafter simply CPL, where $w_0$ and $w_a$ are real numbers. It is usually assumed to parameterize our ignorance about the dynamics of dark energy, and was in particular extensively used by the Dark Energy Task Force \cite{DETF06} as a phenomenological benchmark to compare and contrast the performances of different dark energy probes (see \eg \cite{Virey07}).
Despite its simplicity the CPL parameterization exhibits interesting properties as discussed in detail by Linder \cite{Linder07a}.
In particular,  the two parameters $w_0$ and $w_a$ have a natural physical interpretation: they represent the equation of state's present value and its overall time evolution, respectively. It is argued in \cite{Linderparam,Linder07a} that the best description of $w_a$ in terms of the derivative of $w$ is given by the relation 
\begin{equation}
\label{waderivative}
w_a=-2w'|_{z=1},
\end{equation}
where $w'$ is the logarithmic derivative of $w$ defined as 
$w':=\frac{dw}{d\ln a}$, $a$ being the scale factor of the universe. 

Moreover Linder and Huterer \cite{LinderHuterer05} and Upadhye et al. \cite{Upadhye05} have shown that at most a two-parameter model can optimally be constrained by future data.
Additional major properties are its bounded behaviour 
for high redshift ($\lim_{z \to \infty}w^{CPL}(z)=w_0+w_a =: w_i^{CPL}$) and its ability to describe a large variety
of scalar field dark energy models. Consequently, the CPL parameterization seems to be a good compromise to
define a model independent analysis.\\
Unfortunately, it cannot describe all possible dynamics \cite{Linder07a,WangFreese}, which fact can easily be understood by looking at the dark energy dynamics in the ($w,w'$)-phase space. The CPL parameterization can be re-casted in the following form:
\begin{equation}
\label{lintraj}
w'=-(w_0+w_a)+w,
\end{equation}
that highlights the linear relation between $w$ and $w'$.
A discussion of characteristic phase space properties of several classes of dark energy models can be found in \eg \cite{Linder07a,Caldwell05,CPLphasespace}. It appears that some subclasses are well described by an approximation like eq.(\ref{lintraj}), but that in  general dark energy models do not follow a linear trajectory. Linder \cite{Linderparam} himself for instance argues, that eq.(\ref{CPL}) will hardly be able to handle rapid transitions or oscillations.
\no Consequently there is an unavoidable degree of ``parameterization dependence'' in the results. This rather obvious fact has motivated many different approaches to test the dark energy dynamics. Not only other parameterizations of dark energy's equation of state have been considered 
\cite{eosparam}, but also parameterizations of the dark energy density alone \cite{WangFreese}, or of the Hubble parameter \cite{StarobinskyH}. Finally non-parametric tests have been studied \cite{nonparam,Sahni06}, see \eg \cite{Sahni06} for a general discussion. Since CPL is widely implemented in both observational and theoretical studies, it is essential to test its robustness in reconstructing the dynamics of physically motivated dark energy models.

\no Focusing on quintessence models, Caldwell and Linder \cite{Caldwell05} have shown that the subclasses comprising the so-called ``freezing'' and ``thawing'' models are well parameterized by the CPL functional form, whereas Corasaniti and Copeland \cite{Corasaniti03} pointed out that certain models, in particular the ``tracker models'' \cite{tracker}, would be better described by an equation of state of step-like functional form. Step-like functions are able to describe slow or rapid transitions between two asymptotic values, their modeling is however somewhat arbitrary: a Fermi-function has been used by Bassett et al. \cite{Bassett02}, a linear combination 
of Fermi-functions by Corasaniti and Copeland \cite{Corasaniti03}, a power-law behaviour by Hannestad and M\"ortsell  \cite{Hannestad04},
 an e-fold model by Linder and Huterer \cite{LinderHuterer05}, and hyperbolic tangent functions
by Pogosian et al. \cite{Pogosian2005} and Douspis et al. \cite{Douspis06}. The main drawback of such step-like equations of state is the necessity to introduce four parameters. Given that large number of degrees of freedom, this kind of 
parameterization does therefore not seem to be the appropriate choice to extract constraints from data, even if an interesting and non-trivial study in this direction has been done in \cite{Douspis06}.\\
In this article we however propose to test the significance of the physical information enclosed in the CPL parameterization, obtained from a fiducial cosmology that is described by a step-like model of dark energy.
For if we take seriously the task of testing a possibly wide range of dark energy models 
with future cosmological probes like SNAP/JDEM, we will have to use a 
parameterization of dark energy's equation of state. But if we do not want to take the 
risk of exluding a model on the basis of a parameterization that may not be the appropriate description of the actual dark energy phenomenology, we will have to quantitatively know the intrinsic limitiations of the specific
parameterization we chose.\\
In the following section \ref{Approach} we present and discuss the explicit functional form of our fiducial model, and set up our data framework consisting of supernova Ia data as forecasted for a satellite mission like SNAP/JDEM in combination with future expectations for the CMB shift parameter $R$ and the BAO parameter $A$. After a discussion of our analysis strategy we will, in section \ref{Results}, present our results, and give a summary and discussion in section \ref{Conclusions}. The article closes with an outlook on future prospects.
\section{\label{Approach}Approach}

We choose the hyperbolic tangent functional form first used by Douspis et al. \cite{Douspis06} to model the fiducial step-like dark energy equation of state: \\
\begin{equation}
\label{tanh}
 w^{step}(z)=\frac{1}{2}(w_i+w_f)-\frac{1}{2}(w_i-w_f) \tanh{\left [ \Gamma\ln{\left ( \frac{1+z_t}{1+z} \right )} \right ]},
\end{equation}
where four parameters are introduced: $w_i$ is the equation of state's value at early times: 
$w_i=\lim_{z \to \infty} w^{step}(z)$, $w_f$ its future value: 
$w_f=\lim_{z \to -1} w^{step}(z)$, $z_t$ marks the redshift at the step's center: 
$w^{step}(z_t)=w_{av}:=\frac{1}{2}(w_i+w_f)$, and $\Gamma>0$ rules the width of the 
transition (cf. eq.(\ref{transitionwidth}) below). The advantages of the parameterization given by eq.(\ref{tanh}) are its analytic integrability and the fact, that the equation of state's asymptotic values before and after the transition, $w_i$ and $w_f$, are decoupled. The Hubble function is calculated to be
\begin{eqnarray}
\label{Hubble}
\left (\frac{H(z)}{H_0}\right )^{2}&=&\Omega_M(1+z)^3+\Omega_{DE}(1+z)^{3(1+w_{av})} \nonumber \\
&& \times {\left [ \frac{ {\left ( \frac{1+z_t}{1+z} \right )}^{\Gamma}+{\left ( \frac{1+z_t}{1+z} \right )}^{-\Gamma} }{ {(1+z_t)}^{\Gamma}+{(1+z_t)}^{-\Gamma} } \right ]}^{\frac{3\Delta w}{2\Gamma}},
\end{eqnarray}
where radiation and curvature contributions $\Omega_R$ and $\Omega_K$ are neglected, and $\Delta w:=w_i-w_f$ is the amplitude of the transition. The models described by eq.(\ref{tanh}) represent more general dynamics in the ($w$,$w'$)-phase space than the CPL models,
since their trajectories are parabolae and not simple straight lines any more:
\begin{equation}
\label{partraj}
w'=2\Gamma \left (\frac{(w-w_{av})^2}{\Delta w} - \frac{1}{4}\Delta w \right ).
\end{equation}
Finally, the transition width $\Gamma$ can easily be related to some redshift interval $\Delta z$ around $z_t$. 
The transition from $w_i$ to $w_f$ takes place in the redshift interval
\begin{equation}
\label{transitionwidth}
\Delta z=2 (1+z_t)\sinh(2\,\Gamma^{-1}).
\end{equation}
To derive eq.(\ref{transitionwidth}) we define $\Delta z$ as the interval between the redshifts 
where eq.(\ref{tanh}) takes the values $w_{av} \pm \frac{1}{2}\Delta w\tanh(2)$. Since $\tanh(2)\approx0.96$,
this criterion captures the essence of the dark energy dynamics.
We note, that the redshift width $\Delta z$ decreases with increasing positive $\Gamma$ values, 
but is also linearly dependent on $z_t$. For $z_t=0$ we obtain for example $\Delta z=0.5,1,10$ for $\Gamma=8.08,4.16,0.86$, respectively. \\

For our analysis we use the program \textit{Kosmoshow}. The minimisation procedure is described in \cite{Virey04}. Our dataset consists of simulated data from a future space mission like SNAP/JDEM, that plans to discover around 2000 identified Type Ia Supernovae at redshifts $0.2<z<1.7$ with very precise photometry and spectroscopy. The Supernova distribution is given by \cite{Kim}, see also \cite{Virey04}. We neglect the effect of adding some systematical errors for the magnitude, and we use an additional dataset of 300 nearby Supernovae as expected by the SN Factory \cite{SNFactory}. We combine these simulated data with the CMB shift-parameter $R$ \cite{Rpara} and the BAO parameter $A$ \cite{Eisenstein05}, where we assume an error of $\pm0.007$ on $R$ \cite{Linder07a} (which is the estimate for future PLANCK data \cite{Planck}) and an error of $\pm0.005$ on $A$. These expected errors on magnitudes correspond to a long term scenario (2015-2020), or a Stage IV data model as defined in the report of the Dark Energy Task Force \cite{DETF06}. We neglect the radiation component and will assume spatial flatness in the following. \\
For the fiducial cosmology\footnote{To avoid any confusion between fiducial and fitted quantities, we add a superscript $F$ to the fiducial $\Omega_M$ and $M_s$. This is not done for the fiducial parameters $w_i$, $w_f$, $z_t$ and $\Gamma$, since there is no ambiguity with the fitted $w_0$ and $w_a$ CPL parameters.} we fix $\Omega_M^F$ to 0.3 (hence $\Omega_{DE}^F$ is fixed to 0.7) and the normalisation parameter for SNIa $M_s^F$ to $3.6$.\footnote{$M_s$ is the normalisation constant that enters into the luminosity-distance relation as $m(z)=M_s+5log(\frac{c}{H_0}d_l(z))$.}
To describe dark energy we use the Hubble function of the step-like model given by eq.(\ref{Hubble}). We will consider slow or rapid transitions occurring at low and high redshift for various choices of $w_i$ and $w_f$, precise values will be given in the result section.  
The fiducial cosmological parameters being fixed we are now able to simulate our ``observables'', namely the supernovae magnitudes plus $R$ and $A$.

We then fit the resulting observables with the Hubble function 
\begin{equation}
\left ( \frac{H(z)}{H_0}\right )^2=\Omega_M(1+z)^3+\Omega_{DE}(1+z)^{3(1+w_0+w_a)}e^{-3w_a\frac{z}{1+z}} \nonumber
\end{equation}
calculated from 
eq.(\ref{CPL}). 
We perform fits on the SNIa normalisation parameter $M_s$, the present matter-density fraction $\Omega_M$, 
and the CPL dark energy equation of state parameters $w_0$ and $w_a$. \\

The first information to look at is the value of the $\chi^2$. In real data analysis, a wrong assumption can be detected through a simple $\chi^2$ test:
a high $\chi^2$ indicates that the performance of the fit is bad. This can be the indication of a problem, whose identification is usually not easy in practice. With simulated data, we know the fiducial model and we control the fitting procedure, then a high $\chi^2$ is directly the indication of a wrong assumption in the analysis. We apply as evaluation criterion cuts at $1\sigma$ or $2\sigma$ on the $\chi^2$ values.
The {\it rms} of the $\chi^2$ is $\sigma(\chi^2)=2N_{dof}$, where $N_{dof}$ is the number of degrees of freedom in the fit. If $\chi^2>2N_{dof}$, we consider the wrong assumption to be detected. Conversely, if  $\chi^2<2N_{dof}$, we don't have any indication of something going wrong. In this case, when in addition biases are present, we may misinterpretate the data. $N_{dof}$ will be $16$ in all studies presented in this article.

For the purpose of comparison of the fitted equation of state with the fiducial one we will test the reconstruction of:\\
i) $\Omega_M$ (\ie comparison of $\Omega_M^F$ with $\Omega_M$),\\
ii) the present value of the dark energy equation of state $w(0)$ (\ie $w^{step}(0)$ vs. $w_0$),\\
iii) the value of the dark energy equation of state at the pivot redshift $w(z_p)=:w_p$ where the error on $w(z)$ is the smallest\footnote{see \cite{DETF06,Virey07,Hu} for definitions} (\ie $w_p^{step}$ vs. $w_p^{CPL}$),\\
iv) the overall time evolution of the dark energy equation of state encoded in the $w_a$ parameter along with the relation eq.(\ref{waderivative}) (\ie $-2(w^{step})'|_{z=1}$ vs. $w_a$), \\
v) the initial value of the dark energy equation of state $w_i$ (\ie $w^{step}_i$ vs. $w^{CPL}_i:= w_0+w_a$), to get some insight into the high redshift behaviour.

\no We define the bias of the parameter $p$ by $B_p=|p^F-p|$ and say that $p$ is biased (valid) if the bias is larger (smaller) than the error obtained for $p$, \ie if $B_p>\sigma (p)$ ($B_p<\sigma (p)$).
We also define the Bias Zone (Validity Zone) as the set of all fiducial models 
where the $p$ parameter is biased (valid). Consult \cite{Virey04} for more details
on these definitions.

\no From the comparison of these five quantities we will be able to infer if the CPL parameterization allows a relevant
measurement of the cosmological parameters in case of a rapid transition of the dark energy equation of state.\\

\section{\label{Results}Results}

\subsection{Illustration}

To illustrate the problem we start our discussion with two examples. We define two fiducial models having a fast transition (we fix $\Gamma =10$) that differ by the redshift of their transition: model A's transition is centered at $z_t=3$ (a transition outside the redshift range of SNIa and BAO data) and model B's at $z_t=0.5$ (a recent transition within reach of available data). Motivated by tracker models we fix the remaining two parameters $w_i$ and $w_f$ to $0$ and $-1$, respectively.

\no For both fiducial models we fit the associated cosmology with the CPL parameters $w_0$ and $w_a$ (along with $\Omega_M$ and $M_s$) and get the results shown in Figure 1 and Table 1.
For model A, we obtain a good reconstruction of the local cosmological parameters but a biased estimation of the overall time variation and of the high redshift behavior. Namely, $\Omega_M$, $w_0$ and $w_p$ are valid, but $w_a$ and $w_i$ are biased. For model B all the parameters are biased. We note that in this case $\chi^2=27.5<1\sigma$ clearly indicates a fit of bad quality, but does not yet allow to reject the fit results. 

\begin{figure}[h]
\begin{center}
\includegraphics[width=8cm]{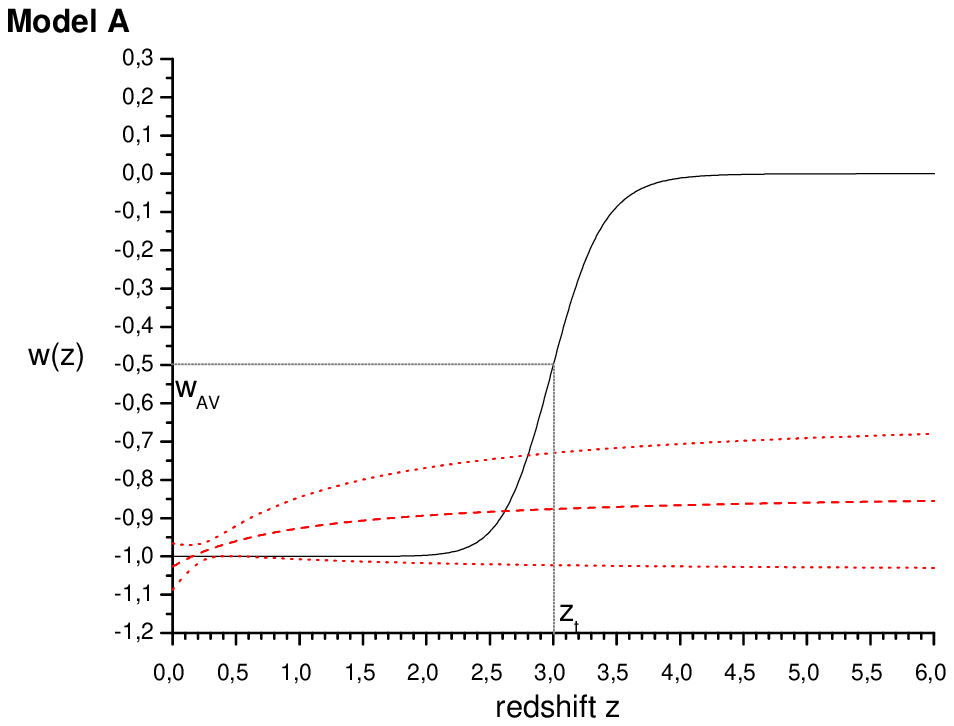}
\includegraphics[width=8cm]{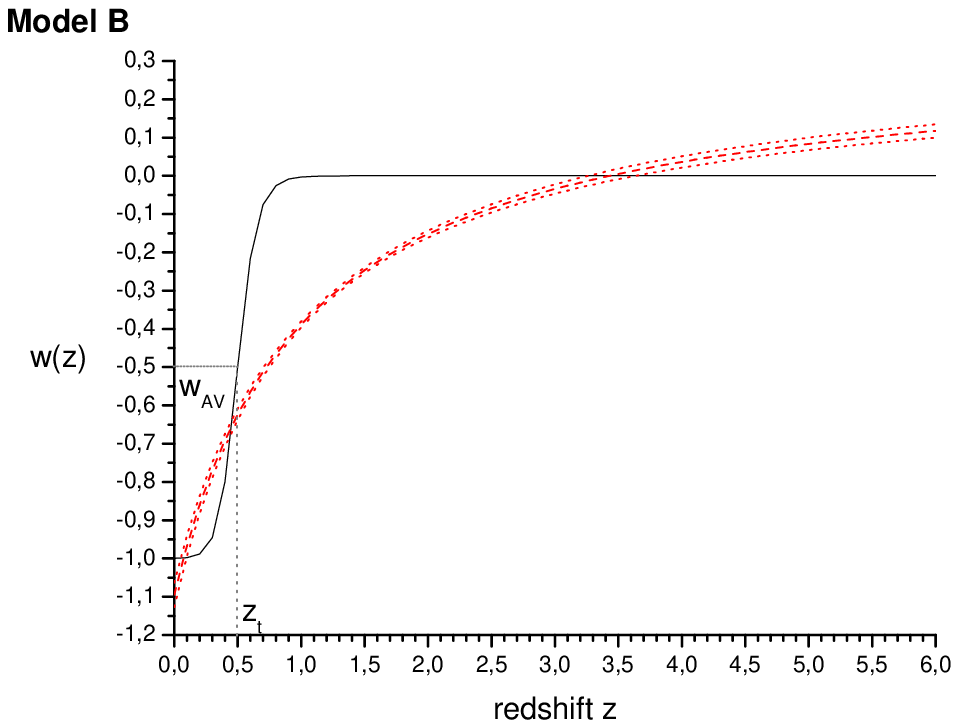}
\caption{
\label{fig:kosmofitsharp}\small{
Results of the fits based on the use of the CPL dark energy equation of state parameterization for two different fiducial step-like dark energy models A (above) and B (below). These fiducial models are such that $\Omega_M^F=0.3$, $w_i^F=0$ and $w_f^F=-1$ with a rapid transition $\Gamma =10$, centered at $z_t=3$ for model A and at $z_t=0.5$ for model B.  The transition width is $\Delta z=1.6$ in case A and $\Delta z=0.6$ in case B. The fiducial step-like dark energy equation of states are plotted in full lines, the reconstructed CPL equation of states are the dashed curves along with the associated $1\sigma$ errors (dotted curves).
}}
\end{center}
\end{figure}

\begin{table}[h!]
\begin{center}
\begin{tabular}{rl|ccc|ccc}
 & &\multicolumn{3}{c|}{Model A} &\multicolumn{3}{c}{Model B} \\
 & & fid & fit & $\sigma$  & fid & fit & $\sigma$  \\
\hline
\hline

i)& $\Omega_M$\ & $0.3$ & $0.295$ &  $0.006$ &{$0.3$} & \underline{$0.314$} & {$0.006$} \\

ii)& $w_0$ & $-1.00$ & $-1.03$ & $0.06$ & {$-1.00$} & \underline{$-1.10$} & {$0.03$} \\

iii) & $w_p$ & $-1.00$ & $-0.99$ & $0.02$ & {$-0.023$} & \underline{$-0.292$} & {$0.007$} \\

iv) & $w_a$ & $0$ & \underline{$0.26$} & $0.24$ & {$0.13$} & \underline{$1.41$} & {$0.0.05$} \\

v)& $w_i$ & {$0$} & \underline{$-0.86$} & {$0.21$} & {$0$} &\underline{$+0.32$} & {$0.02$} \\

& $\chi^2$ &  & $0.9 $ &  &  & 27.5 & \\

\end{tabular}
\caption{\label{table}\small{Fiducial and fitted values of the five parameters of study
(see section 2)
for models A and B. The pivot redshift is $z_p=0.27$ ($z_p=1.31$) for model A
(model B). Biased fitted values are underlined.}}
\end{center}
\end{table}

\no From these two examples we find that the CPL parameterization should allow a valid reconstruction of the local (\ie present value) cosmological parameters even in case of a rapid transition if the transition is not recent. Conversely and without surprise, the high redshift behavior can lead to misinterpretations. In the following we vary all four fiducial parameters to test the stability of these results.\\

\subsection{\label{Sec:ztGammaPlane}The ($z_t$,$\Gamma$)-plane}

\begin{figure}[h]
\begin{center}
\includegraphics[width=8cm]{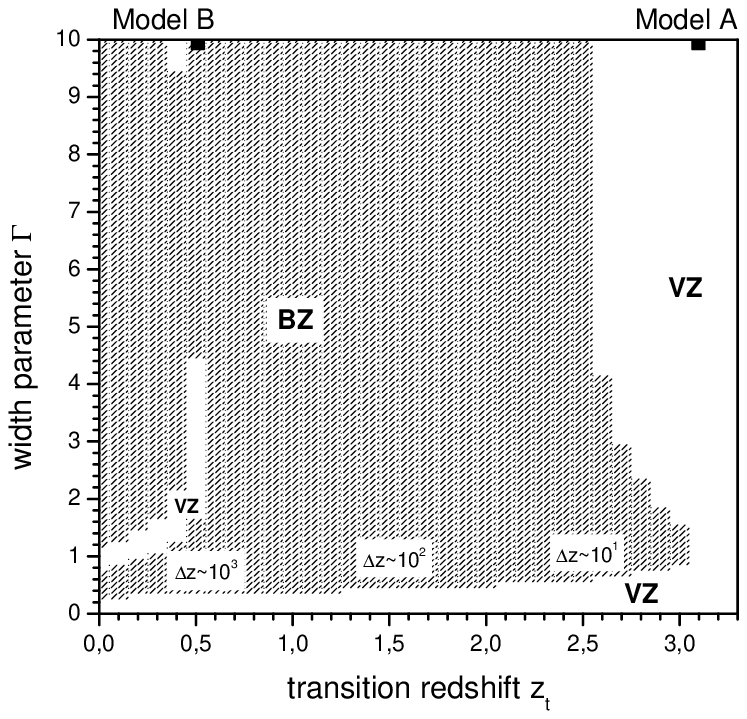}
\includegraphics[width=8cm]{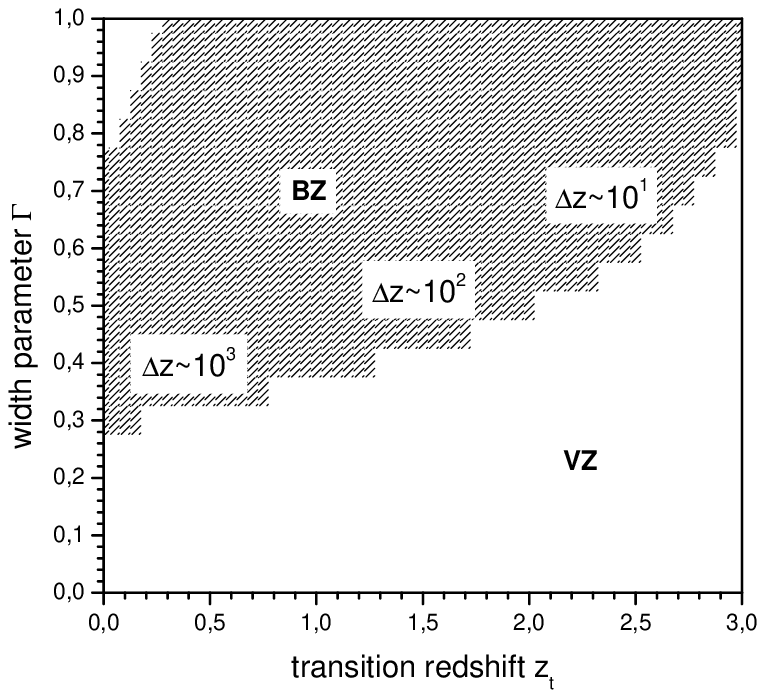}
\caption{\label{fig:ztGammaevolutionw0}\small{Quality of reconstruction of $w_0$ for fiducial
step like models for dark energy in the ($z_t$,$\Gamma$) parameter space, where $w_i=0$ and $w_f=-1$. We give the Bias Zone (BZ, hatched) and the Validity Zone (VZ, white) obtained for the CPL parameter $w_0$ in the full plane (above) and in higher resolution in the $(z_t,\Gamma)=([0;3],[0;1])$-section of the plane (below). We also marked the order of magnitude of $\Delta z$  along the border of the VZ.}}
\end{center}
\end{figure}

\begin{figure}[h]
\begin{center}
\includegraphics[width=8cm]{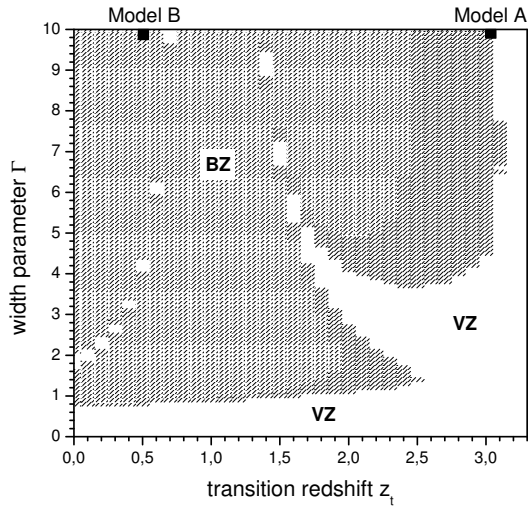}
\caption{\label{fig:ztGammaevolutionwa}\small{Quality of reconstruction of $w_a=-2w'|_{z=1}$ for fiducial
step like models for dark energy in the ($z_t$,$\Gamma$) parameter space, where $w_i=0$ and $w_f=-1$. We give the Bias Zone (BZ, hatched) and the Validity Zone (VZ, white) obtained for the CPL parameter $w_a$.}}
\end{center}
\end{figure}

If we henceforth keep $w_i$ and $w_f$ fixed to their values $0$ and $-1$, respectively, we can study our step-like models in a two parameter phase space: the ($z_t$,$\Gamma$)-plane. In this plane each point represents a fiducial model, and for each one we perform a fit with the CPL equation of state. Then we test the $\chi^2$ value and the reconstruction of the five parameters of study. Our results are given in Figures \ref{fig:ztGammaevolutionw0} and \ref{fig:ztGammaevolutionwa}, where we chose $z_t$ in the range $[0;3.3]$ and $\Gamma$ in $[0;10]$. We however performed a complete scan up to $z_t=5$ and checked the stability of our results for even higher values of $z_t$ and $\Gamma$.
It appears that the $\chi^2$ is below $1\sigma$ for all the models in the presented plane, except for a small region at $0.2<z_t<0.4$ and $\Gamma>6$, where $\chi^2>2N_{dof}$ (but $\chi^2<4N_{dof}$). The fit quality therefore being sufficiently good, it will be of crucial interest to study the quality of reconstruction of the cosmological parameters. Figure \ref{fig:ztGammaevolutionw0} shows the quality of the reconstruction of $w_0$ in this plane, where we give the Bias and Validity Zones obtained for $w_0$.

We recover the previously introduced models A and B, and make the following remarks: \\
i) $w_0$ (as well as all other parameters) is fully reconstructed along the line $\Gamma=0$. This is merely a sign of consistency, since $\Gamma=0$ imposes $w_i=w_f$ on the fiducial model, which in turn simply means a $w=const.$ behaviour that must be well reconstructed by CPL. \\
ii) The lower limit of the Validity Zone gives the bounds on the width the transition where the use of CPL equation of state is still justified. We see that if $\Gamma \lesssim 0.3$, then the transition is sufficiently slow for the CPL parameterization to be a reasonable description of the dark energy dynamics \textit{whatever} $z_t$. The limit is slowly increasing to higher $\Gamma$-values with increasing $z_t$, forming a concave curve, but does \textit{not} follow a line $\Delta z=const$. It reaches $\Gamma=0.8$ at $z_t=3$, where $\Delta z \approx 50$. We marked the order of magnitude of $\Delta z$  along the border of the VZ in Figure \ref{fig:ztGammaevolutionw0}. 
Note however that these bound depends on the values of $w_i$ and $w_f$, which will be the point of discussion in the next paragraph.\\
iii) We find a good reconstruction of $w_0$ for all $z_t \gtrsim 2.5$ whatever $\Gamma$ is, except a small bulge between $\Gamma$-values $[0.6;2.2]$, that extends (exactly) up to $z_t=3.0$. 
This can easily be understood thanks to Figure \ref{fig:kosmofitsharp}, where we see that for high-redshift-transitions ($z_t=3$ for model A) the low redshift behaviour mimics a constant-$w$-model in case of a fast transition, or a "nearly" constant-$w$-model in case of a slow transition. Both those behaviours are well reconstructed by the CPL equation of state.\\
iv) We discover a little zone of good reconstruction for $z_t=0.5\pm0.1$, when $1.8\lesssim \Gamma\lesssim4.6$ (which corresponds to $1.2 \lesssim \Delta z \lesssim 3.5$). However we did not find a compelling physical argument for its appearance, indicating an accidental valid reconstruction. Consequently, this small Validity Zone is not particularly interesting.\\

The scan of the same ($z_t$,$\Gamma$)-plane for $w_p$ leads to qualitatively similar results.
We note however that the Bias Zone is enlarged compared to the one for $w_0$: $w_p$
reconstruction is valid if $z_t>3$ whatever $\Gamma $, or
if $\Gamma<0.2$ whatever $z_t$. We note however, that the reconstruction of $w_p$ is highly sensitive to the errors ascribed to the supernovae magnitudes and the parameters $R$ and $A$, and we do therefore not consider the pivot redshift as a good mean of interpretation of our fit, in agreement with \cite{Linder2006}. 

\no For $\Omega_M$ we find results  similar to those for $w_0$, with a small increase of the Bias Zone in the $z_t$-direction:
the $z_t$ limit is at $z_t\approx3.2$, and a decrease of the Bias Zone in the $\Gamma$-direction, where the limit is now located at $\Gamma\approx0.6$.

The overall time evolution of the dark energy equation of state is encoded in the $w_a$ parameter and the relation eq.(\ref{waderivative})
has been proposed for its concrete interpretation. In our approach we test this relation through a comparison of the fitted $w_a$ with $-2w'|_{(z=1)}$ calculated for the fiducial step-like model. Our results are given on Figure \ref{fig:ztGammaevolutionwa}. Eq.(\ref{waderivative}) allows a
correct interpretation of $w_a$ if: i) the transition width $\Gamma<0.8$ whatever $z_t$ (\ie $\Delta z>12$), ii) the transition center $z_t>3.1$, whatever $\Gamma$, iii) and in the range $2.5<z_t<3.1$ if $\Gamma<4$ (\ie $\Delta z>4$). Outside these domains, the validity of eq.(\ref{waderivative}) breaks down and we loose the meaningful physical interpretation of the $w_a$ parameter.

Concerning the reconstruction of the high redshift behaviour, it appears that
$w_i$ is very badly reconstructed for nearly all pairs ($z_t$,$\Gamma$), where $\Gamma>0.1$.
In fact, a Validity Zone exists ($1.3<z_t<1.5$ and $2<\Gamma <10$), but it is accidental and its
position changes a lot if we change the $w_i$ and/or $w_f$ parameter. We consequently have strong chances to misinterpret the high redshift behaviour of the dark energy equation of state when using the CPL parameterization.\\
\no We note here that we also checked the reconstruction of the ``CMB effective value'' of dark energy's equation of state that was proposed by Huey et al. \cite{Huey1999}, and also studied by Pogosian et al. \cite{Pogosian2005}: 
\begin{equation}
w_{eff} := {\int_{0}^{z_{cmb}}w_x(z)\Omega_{DE}(z)dz}/{\int_{0}^{z_{cmb}}\Omega_{DE}(z)dz}, \nonumber
\end{equation}
where $z_{cmb}=1089$ from \cite{WMAP}. It's reconstruction however shows up to be as problematic as the one of $w_i$.\\

Consequently, from Figures \ref{fig:ztGammaevolutionw0} and \ref{fig:ztGammaevolutionwa} we have been able to quantify a validity range
of the CPL parameterization in terms of the position of the transition ($z_t\gtrsim 2.5$ whatever
its rapidity) or in terms of the width of the transition ($\Gamma \lesssim 0.3$ whatever $z_t$)
for the tracker models pointing to a cosmological constant in the future (\ie $w_i=0$ and
$w_f=-1$). Unfortunately these bounds strongly depend on the $w_i$ and $w_f$ parameters.
For example, if we change the parameter $w_i=0$ to $w_i=-0.8$, keeping $w_f$ fixed to $-1$, the $\chi^2$ comes out to be extremely low ($\chi^2<0.15$) in the whole plane, and the Validity and Bias Zones in the ($z_t$,$\Gamma$)-plane change a lot. Now, $\Omega_M$ is always valid whatever $z_t$ and $\Gamma$ are. $w_0$ is biased only if $z_t<0.25$ and $\Gamma > 7$ (\ie $\Delta z<0.7$), namely for very rapid and very recent transitions. The interpretation of $w_a$ is biased if $z_t<1.5$ with $\Gamma>2$ (\ie $\Delta z<5.7$).  Surprisingly, $w_i$ is biased only if $z_t \approx 0.3\pm 0.1$ and $\Gamma > 6$ (\ie $\Delta z<0.9$). This means that CPL is able to catch the high redshift behaviour of the dark energy dynamics if $z_t>0.4$ whatever the width of transition.
For this particular example we hence conclude that CPL is an extremely good choice of
parameterization for the dark energy equation of state.\\

\subsection{The ($w_i$,$w_f$)-plane}

\begin{figure}[h!]
\begin{center}
\includegraphics[width=8cm]{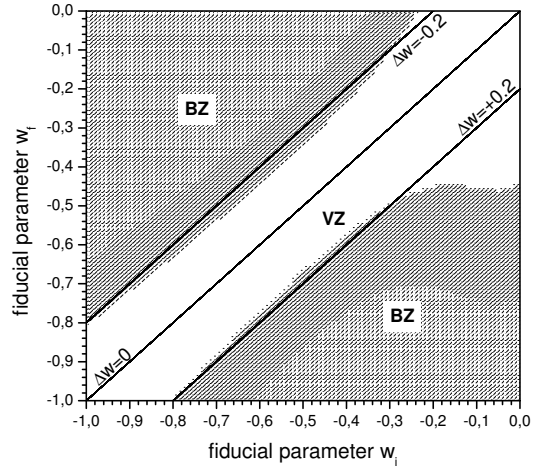}
\caption{
\label{fig:wiwfw0} \small{
Quality of reconstruction of the $w_0$ CPL parameter for fiducial dark energy step-like models in the ($w_i$,$w_f$) parameter space with $z_t=0.5$ and $\Gamma=10$ (\textit{i.e.} $\Delta z = 0.6$). We give the Bias Zone (BZ, hatched) and the Validity Zone (VZ, white) obtained for $w_0$. We also plot some lines of constant transition amplitude $\Delta w$. }
}
\end{center}
\end{figure}

\begin{figure}[h!]
\begin{center}
\includegraphics[width=8cm]{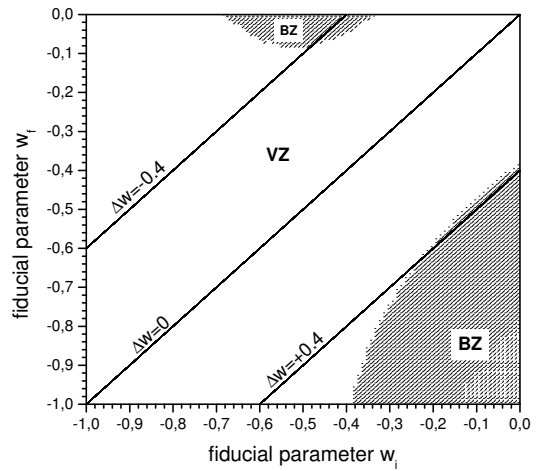}
\caption{
\label{fig:wiwfom} \small{Same as Figure \ref{fig:wiwfw0} but for parameter $\Omega_M$ obtained with the CPL equation of state.}
}
\end{center}
\end{figure}

\begin{figure}[h!]
\begin{center}
\includegraphics[width=8cm]{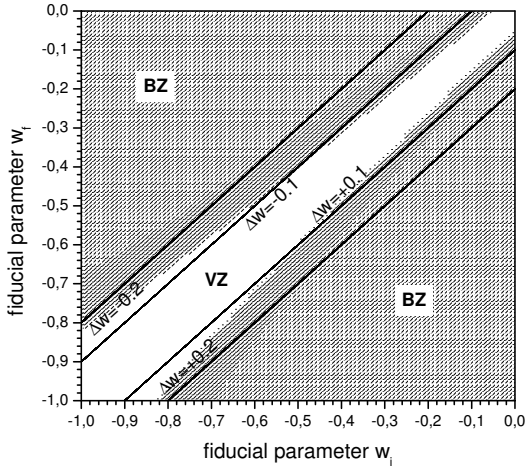}
\caption{
\label{fig:wiwfwi} \small{Same as Figure \ref{fig:wiwfw0} but for the $w_i:=w_0+w_a$ parameter obtained with the CPL equation of state. A very similar figure with a slightly narrower validity zone is obtained for the reconstruction of the $w_a$ parameter through eq.(\ref{waderivative}).}
}
\end{center}
\end{figure}

\no To be more quantitative on the effect of the variations of the $w_i$ and $w_f$ parameters,
we study the biases in the $(w_i,w_f)$-plane for \textit{the most pessimistic case} for
$z_t$ and $\Gamma$: we fix $z_t=0.5$ and $\Gamma =10$ (\ie $\Delta z=0.6$). When $w_i=0$ and $w_f=-1$ this corresponds to our model B where all fitted parameters where biased (cf. Table \ref{table} and Figure \ref{fig:kosmofitsharp}). We consider variations
for $w_i$ and $w_f$ in the range $[-1;0]$ for both. Figures \ref{fig:wiwfw0}, \ref{fig:wiwfom} and \ref{fig:wiwfwi} show the Validity and Bias Zones in the ($w_i$,$w_f$)-plane for the parameters $w_0$, $\Omega_M$, and $w_i$. We get:\\
i) $w_0$ is valid if $|\Delta w|\lesssim 0.2$.\\
ii) $\Omega_M$ is well reconstructed whatever $w_f$ is, if $w_i\lesssim -0.7$.
If $w_i\gtrsim-0.4$, $\Omega_M$ is valid only if $|\Delta w|=|w_f-w_i|\lesssim 0.4$.\\
iii) $w_i$ is valid if $|\Delta w|\lesssim 0.1$, which limit increases to $|\Delta w|\lesssim 0.2$
when $w_i\lesssim-0.8$. Similarly, we find that the reconstruction of the $w_a$ paremeter through eq.(\ref{waderivative}) is valid only if $|\Delta w|<0.1$.\\
We note that for our choice of $z_t$ and $\Gamma$ we find that $w_p$ is valid if $|\Delta w|\lesssim 0.1$ when both $w_i$ and $w_f$ are smaller than $-\frac{1}{3}$. Hence, $w_p$ is more likely to be biased than $w_0$. This weakens the usefulness of $w_p$, as was already inferred in Section \ref{Sec:ztGammaPlane}.

We consequently find that the CPL parameterization is able to yield valid results for the
cosmological parameters even for a very fast and recent transition (the worst situation),
if and only if the transition amplitude $\Delta w$ is not too large.\\

\subsection{Confusion with $\Lambda$CDM and $w=const.$ models}

As soon as biases are present in the analysis it is interesting to study the actual values of the biased parameters, in order to know if we can confuse the true cosmology with a simpler model, such as the $\Lambda$CDM or more generally the models with constant equation of state $w$. We performed this exercise and found that if $z_t\gtrsim3$ and $\Gamma\gtrsim2$ we confuse the true cosmology with a
$w=const.$ model. If, in addition, $w_f=-1$ then the confusion is with $\Lambda$CDM. This can easily be understood  from the Model A plot of Figure 1: the true cosmology effectively corresponds, at low redshifts where SNIa and BAO data are located, to a $w=const.=-1$ model. The high redshift behavior is only weakly constrained by the CMB. This is in agreement with the calculations performed in \cite{Douspis06}, which show that for a $\Lambda$CDM fiducial cosmology one gets almost no constraint on the transition width $\Gamma$ when the location of the transition is bigger than $z_t>0.8$ \cite{Douspis06}. For other values of $z_t$ and $\Gamma$ such confusions seem impossible (except for exotic phantom models having both $w_i$ and $w_f$ below $-1$).


\section{\label{Conclusions}Conclusions}

We quantified the degree to which CPL's parameterization of dark energy's equation of state 
eq.(\ref{CPL}) would be able to cover a rapid, step-like time evolution of dark energy's equation of state. We used a hyperbolic tangent function to model such a step, and performed the fit of Supernova Ia data from a future space mission like SNAP/JDEM in combination with future expectations for the CMB shift parameter $R$ and the BAO parameter $A$.\\
We found that the cosmological parameters describing the recent expansion of the universe,
namely the matter density $\Omega_M$ and the present value of the dark energy equation of state $w_0$ are well reconstructed \textit{except} for a recent $z_t\lesssim2.5$ and rapid $\Gamma\gtrsim0.3$ transition with a large amplitude $|\Delta w|=|w_i-w_f| \gtrsim0.4$. 
The value at the pivot redshift $w_p$ has stronger chance to be biased than $w_0$. Since our results
are rather unstable and the pivot redshift $z_p$ has no physical meaning, we conclude 
that $w_p$ is not a good mean to interprete the data. 
The overall time evolution of the dark energy equation of state, encoded in the $w_a$ parameter via eq.(\ref{waderivative}), is surprisingly well reconstructed. We find that biases are present only if $z_t<3$ and $\Gamma>0.8$ for large amplitudes $|\Delta w|$, and that these bounds are reduced to $z_t<1.5$ and $\Gamma>2$ for small amplitudes. When the amplitude $|\Delta w|$ is smaller than $0.1$, we find no bias at all. \\
Conversely, the high redshift behaviour of the dark energy equation of state is in general strongly biased. It is only in the case of a very slow transition, $\Gamma\lesssim0.1$, or with small amplitudes, $|\Delta w|<0.1-0.2$, that the correct dynamics are obtained. The parameters which have the best reconstruction are thus $\Omega_M$ and $w_0$. This can easily be understood, since the other parameters ($w_p$, $w_a$ and $w_i$) are dependent on the $w_a$ parameter, which has a valid reconstruction only in a smaller parameter space. It appears that it is essentially the high redshift behaviour of the dark energy equation of state, through the $w_i$ parameter, that carries the largest risk of misinterpretation. We should therefore be careful with the interpretation of the initial value of the dark energy equation of state obtained with the CPL parameterization $w^{CPL}(z\rightarrow\infty)=w_0+w_a$. We see from our results that early dark energy models having a sizeable density at recombination and tracker models, will, when we use the CPL parametrization, be confused with a Cosmological Constant if the transition from $0$ to $-1$ is fast and beyond $z=3$.\\

Our results confirm that the CPL parameterization has the quality to catch the dynamics of many dark energy models, and in particular the dynamics of step-like ones. Only for a recent and rapid transition in the dark energy equation of state with a large amplitude the CPL parameterization breaks down in an undetectable way. To rule out such a possibility it will be necessary to perform the cosmological analyses also with a step-like parameterization, like in \cite{Douspis06}. The four parameter phase space of the step-like parameterization should be restricted to the domain where CPL breaks down, namely where $z_t<2.5$ and $\Gamma>0.3$ and $|\Delta w|>0.4$.\\
We recall, that our analysis has been performed in the framework of supernova Ia studies, complemented by distance information from CMB and BAO. The combination with other cosmological probes will certainly have an impact on our conclusions, and will modify the limits of validity that were presented in section \ref{Results}. We however expect our main conclusion to remain valid, namely that a sharp and recent transition of the dark energy equation of state should be explicitely constrained. In this case it could be very interesting to add data that are more sensitive to dark energy dynamics at high-redshift than supernovae and BAO are, such as the Integrated Sachs-Wolfe effect as was considered by \cite{Pogosian2005}.

\section*{Acknowledgements}
We thank Christian Marinoni and Pierre Taxil for fruitful discussions, and Andr\'e Tilquin, who wrote \textit{Kosmoshow} and provided helpful technical support. The work of Sebastian Linden is financially supported by the ``Gottlieb Daimler- und Karl Benz-Stiftung''.

\end{document}